\def\be{\begin{equation}}
\def\ee{\end{equation}}
\def\ber{\begin{eqnarray}}
\def\eer{\end{eqnarray}}
\def\bers{\begin{eqnarray*}}
\def\eers{\end{eqnarray*}}

\def\PR{{ Phys. Rev.}\ }
\def\PRL{{ Phys. Rev. Lett.}\ }

\documentclass[aps,prb,twocolumn,groupedaddress,showpacs,amsmath,amssymb]{revtex4-1}
\usepackage{dcolumn}   
\usepackage{amsmath}
\usepackage{graphicx}    
\usepackage{subfigure}
\usepackage{color}
\newcommand{\comment}[1]{}
\usepackage{ifthen}
\newboolean{includefigs}
\setboolean{includefigs}{true}      
\newboolean{includetext}
\setboolean{includetext}{true}     
\newcommand{\condcomment}[2]{\ifthenelse{#1}{#2}{}}
\begin{document}

\title{Mixed valency and site-preference chemistry for Cerium and its compounds: \\ a predictive DFT study}

\author{Aftab Alam$^{1,2}$ and D D Johnson$^{1,3}$ }
\email[emails: ]{aftab@ameslab.gov (aftab@phy.iitb.ac.in), ddj@ameslab.gov}
\affiliation{$^{1}$The Ames Laboratory, U.S. Department of Energy, Ames, Iowa, USA 50011-3020;}
\affiliation{$^{2}$Physics, Indian Institute of Technology -- Bombay, Powai, Maharastra, Mumbai, India 400076;}
\affiliation{$^{3}$Materials Science \& Engineering, Iowa State University, Ames, Iowa, USA 50011-2300.}

\begin{abstract}
Cerium and its technologically relevant compounds are examples of anomalous mixed valency, originating from  two competing oxidation states -- itinerant Ce$^{3+}$ and localized Ce$^{4+}$. 
Under applied stress, anomalous transitions are observed but not well understood.
Here we treat mixed valency as an ``alloy'' involving two valences with competing and numerous site-occupancy configurations, and  we use density functional theory with Hubbard U ($i.e.$, DFT+U) to evaluate the effective valence and predict properties, including controlling valence by pseudo-ternary alloying. 
For Ce and its compounds, such as (Ce-La)$_2$(Fe-Co)$_{14}$B permanent magnets, we find a stable mixed-valent $\alpha$-state near the spectroscopic value of $\nu_s=3.53$.
Ce valency in compounds depends on its steric volume and local chemistry; for La doping, Ce-valency shifts towards $\gamma$-like Ce$^{3+}$, as expected from steric volume; for Co doping, valency depends on local Ce-site chemistry and steric volume. 
Our approach captures the key origins of anomalous valency and site-preference chemistry in complex compounds.
\end{abstract} 
\date{\today}
\maketitle

{\par} Mixed valence compounds exhibit interesting anomalies when external parameters, such as pressure, are varied. Reliably predicting their properties and determining the origin of mixed valence effects remains open, and dependent upon correlating the properties with the molecular and crystal structure, and with local chemistry.
Apart from the field of metallurgy and pigments in artwork and ceramics (e.g., Fe$_2$O$_3$ and Fe$_3$O$_4$),\cite{Day2008} it also serves as an active area of research in complex biophysical problems, such as photosynthesis, and in organic-conjugated materials\cite{Ang2012} for artificial electronic devices.\cite{CSR12}  
What makes these compounds different from other materials is the coexistence of wide {\it s-d} bands and very heavy atomic-like {\it f}-electrons at/near the Fermi energy. 
Anomalous properties then arise from  a competition between the itinerant and localized nature of the {\it f}- electrons. 
Numerous experiments  have revealed a growing list of such compounds from the rare-earth (RE) series, the late actinides, and some transition-metal compounds. 

{\par} 
Cerium is the first RE element that exhibits phases with enormous ($15-17\%$) volume differences.\cite{phasediag} 
As the most abundant RE, Ce is often considered as a RE-replacement in permanent magnets.\cite{Alam13} 
The low-pressure $\gamma$-phase of Ce exhibits a local magnetic moment, associated with a trivalent Ce$^{3+}$ configuration. With applied pressure, Ce first transforms into a mixed-valent $\alpha$-state with quenched moments, and eventually to a tetravalent Ce$^{4+}$ $\alpha'$-state at higher pressure. 
With alloying a local ``chemical'' pressure can be exerted, so the Ce valency in its compounds can be more sensitive, e.g., Ce$_2$Fe$_{14}$B,  where it remains in a strongly mixed-valent $\alpha$-like state, which is incompatible with a local $4f$-moment.
The central problem is to understand the mechanism that controls Ce valency both pure Ce and its compounds.

{\par} Here, we treat mixed valency by mapping to an ``alloy'' problem, where the Ce$^{3+}$ and Ce$^{4+}$ are considered as two different atoms, and we include the binomial distribution of RE-site occupancy in the lattice. 
Our approach captures the key electronic and chemical effects, reproducing the observed mixed valency of Ce and its complex compounds, such as (Ce-La)$_2$(Fe-Co)$_{14}$B magnets. 
We show that the Ce valency in compounds depends on steric volume of Ce sites and local chemistry surrounding the RE site. 
Mixed valency of Ce is then predicted similar to studies on rare-earth systems using model Hamiltonians,\cite{Ghatak1976} and consistent with a correlated and multi-electron picture of Ce with semi-isolated 4$f$ states in contact with a bath of $spd$ valence electrons, as found experimentally.\cite{EPL2007}

{\par}Addressing mixed valency using a density functional theory (DFT) treats magnetism, atomic multiplet effects, and crystal field splitting on an equal footing, and identifies the electronic mechanisms responsible for the anomalous valence behavior.
While a first-principles Dynamical Mean-Field theory (DMFT) may better describe the fluctuating mix valency, our approach captures the key effects in complex compounds with dramatically less computational intensity. 
Notably, within DMFT $\delta$-Pu is found to be a superposition of two atomic valences (60\% $f^5$ and 40\% $f^6$).\cite{Shim2007} Yet, experimentally, $\alpha$ and $\delta$ Pu have a superposition of three 5$f$ states\cite{PNAS2012} ($\sim$ 20\% $f^4$, 40\% $f^5$ and 40\% $f^6$), a ternary ``alloy'' (2 independent fractions).

{\par} We use Vienna {\it ab-initio} simulation package  VASP\cite{vasp96} with a pseudopotential and projected-augmented-wave basis\cite{vasp99} using Perdew-Burke-Ernzhorf (PBE) exchange correlation and spin-orbit coupling. 
With different sized Ce$^{3+}$ and Ce$^{4+}$ ions, relaxations -- ignored in previous studies -- are crucial to predict reliable energetics and groundstates. 
Localized Ce$^{3+}$ {\it f}-electrons are addressed via a PBE+U approach\cite{PBE+U} with a Hubbard U (set to 5~$e$V from previous work\cite{Alam13}) introduced in a screened Hartree-Fock manner. 
See footnote for more details.\cite{note1}

\begin{figure}[t]
\centering
\includegraphics[width=6.5cm]{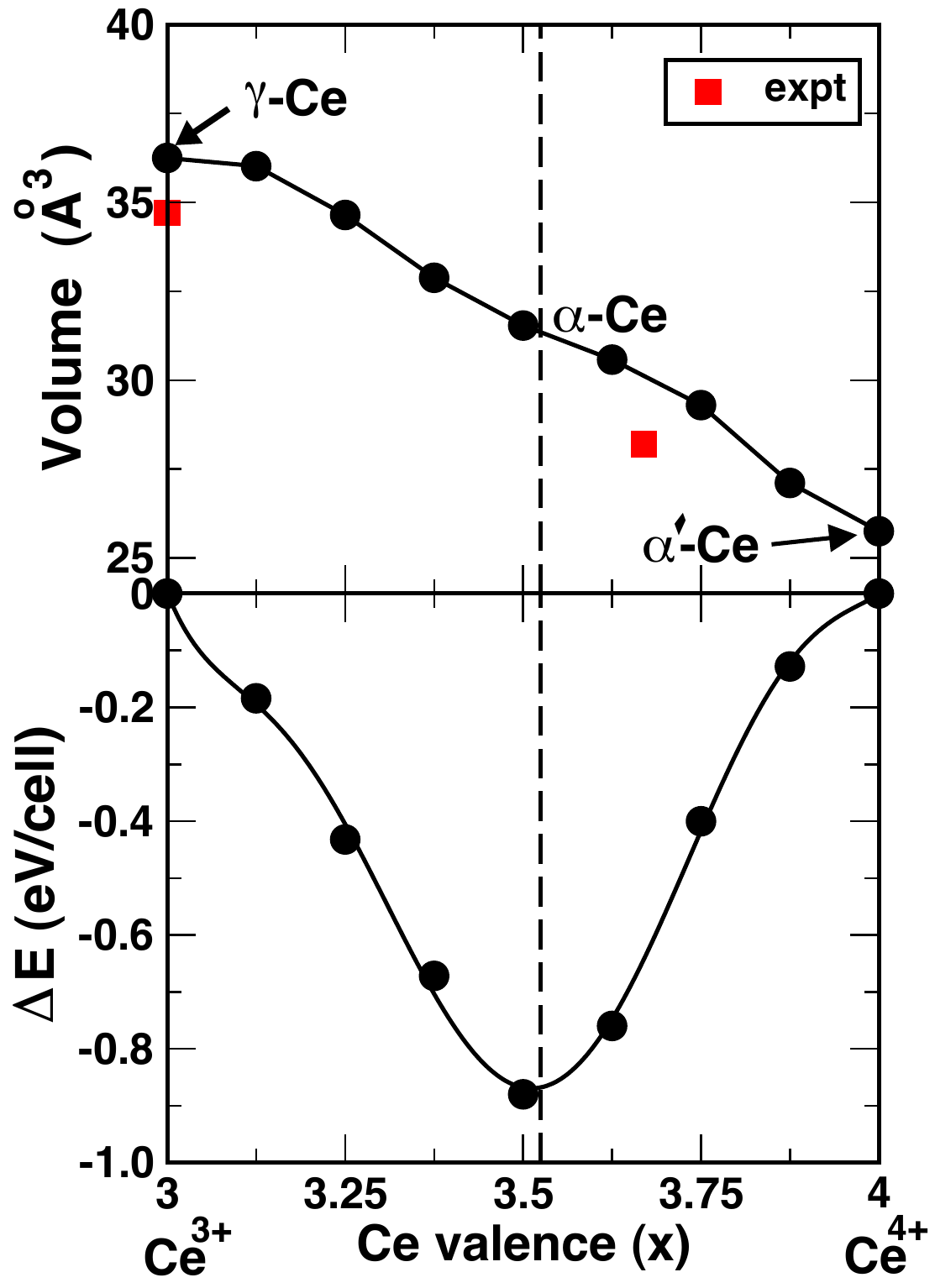}
\caption {(Color online) (Top) Relaxed volume (Bottom) formation enthalpy vs. Ce valence for pure fcc Ce where trivalent (${3+}$) and tetravalent (${4+}$) Ce are mixed. Square symbols indicate experimental volume (from Ref. \onlinecite{Johansson95}). Vertical dashed line is just a guide to the eye for the estimated minimum. 
} 
\label{fig1}
\end{figure}

{\par} The formation enthalpy $\Delta$E and volume V versus Ce-valence in fcc Ce mixed-valence ``alloys'' are shown in Fig.~\ref{fig1}. 
Here Ce$^{3+}$ and Ce$^{4+}$ potentials are occupied over $8$ sites in a Ce supercell to find the energetically most favorable configuration; hence, we have discrete jumps of $0.125$ valence ``composition''. 
From Fig.~\ref{fig1}, the energetically most favorable mixed-valence state occurs near $3.5$ (a $5^{th}$-order polynomial fit yields $\nu_s = 3.55$), near the assessed value of 3.67.\cite{phasediag} 
We find that the mixed valence $\alpha$-state of Ce arises from a energetically favorable distribution of the two Ce$^{3+}$ and Ce$^{4+}$ states. (Using more sites fills in the curve, needed in skewed distributions.)
Atomic positions and the cell volumes are fully relaxed in each data. 
The relaxed, DFT+U calculated V's are compared with known experimental volumes\cite{Johansson95} (red squares), and are within $8\%$ and have the correct trend for the $\gamma$, $\alpha$ and $\alpha'$ phases. 
For pure Ce, the mixed-valency  corresponds to a volume between those of purely $\gamma$- and $\alpha$-phase. 
Including the on-site  U for Ce$^{3+}$ (with $4${\it f}-electron)  and the spin-orbit coupling is important to get the correct groundstate for the intermediate valency.

\begin{figure}[t]
\centering
\includegraphics[width=7.5cm]{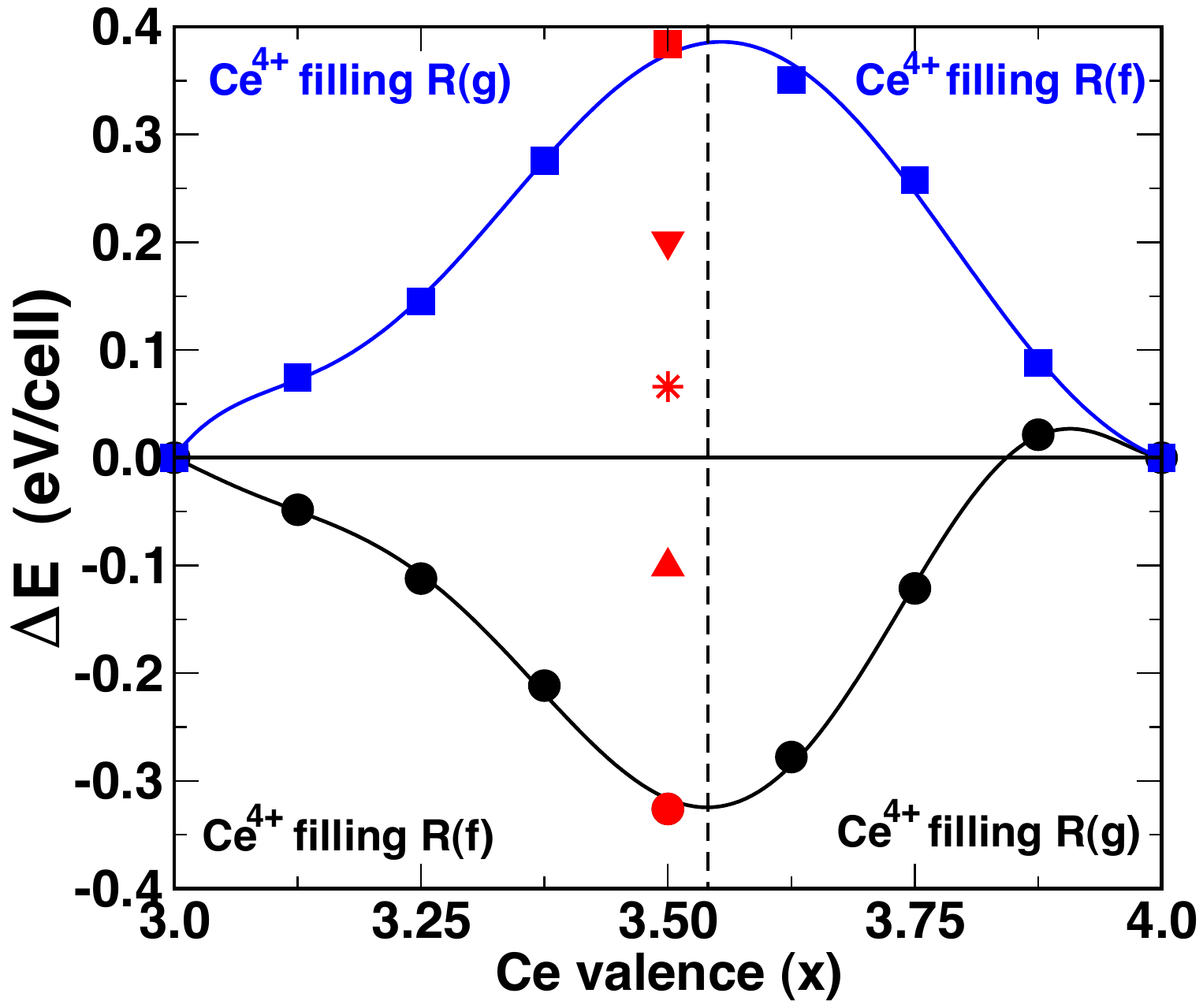}
\caption {(Color online) Mixed valency of Ce in Ce$_2$Fe$_{14}$B. Black (blue) curve 
indicates the formation enthalpies of mixing Ce$^{3+}$ and Ce$^{4+}$, with the larger
$3+$ ions distributed over $4$g ($4$f) sites and smaller $4+$ ions over $4$f ($4$g)
sites. Star and triangles indicate the energies for intermediate sets of distribution where
$3+$ and $4+$ ions are mixed over both the sites. } 
\label{fig2}
\end{figure}

{\par} Next, we investigate the Ce mixed valency in Ce$_2$Fe$_{14}$B, a challenging 68-atom per cell, tetragonal structure with space group P$4_2$/mnm.\cite{Herbst91} 
Mixed valency of Ce significantly affects its magnetic behavior. 
Due to the complex nature of 2-14-1 structure,\cite{Alam13} the Ce mixed valency is associated with Ce site preferences. 
The 2-14-1 structure contains two inequivalent rare-earth (R) sites -- R($4$f) and R($4$g) -- each with multiplicity $4$.\cite{Alam13}
From the coordination shell around each site, $4$g-sites acquire a larger volume than $4$f-sites. 
As such Ce$^{3+}$ (larger ion) prefers to occupy the $4$g-sites while Ce$^{4+}$ (smaller ion) the $4$f-sites. 
Energetically favored configurations are found by mixing Ce$^{3+}$ and Ce$^{4+}$ on the $8$ R-sites in all possible ways.

{\par}The formation energy gain/loss $\Delta$E versus Ce valence in (Ce$^{3+}$-Ce$^{4+}$)$_2$Fe$_{14}$B for all 8 R-site configurations are shown in Fig.~\ref{fig2}. 
The filled circles indicate energies when Ce$^{3+}$ is favorably distributed over $4$g-sites and Ce$^{4+}$ on $4$f-sites; filled squares are the results with the opposite (unfavorable) distribution of Ce-ions. 
Other symbols indicate intermediate sets of distribution where ${3+}$ and $4+$ ions are mixed over both the sites with a binomial distribution.
Notice the asymmetric nature of the energy curves comparing the lower vs. upper half of $\Delta$E, which is due to a different filling on the $2$ inequivalent RE-sites; that is, the collective effect of filling Ce$^{3+}$ ions preferentially over $4$f-sites are very different from that of $4$g-sites. 
The favorable mixed-valency occurs near $\nu_s=3.5$ ($3.55$ from polynomial fit), near the assessed $3.44$.\cite{Capehart93}

\begin{figure}[t]
\centering
\includegraphics[width=7.5cm]{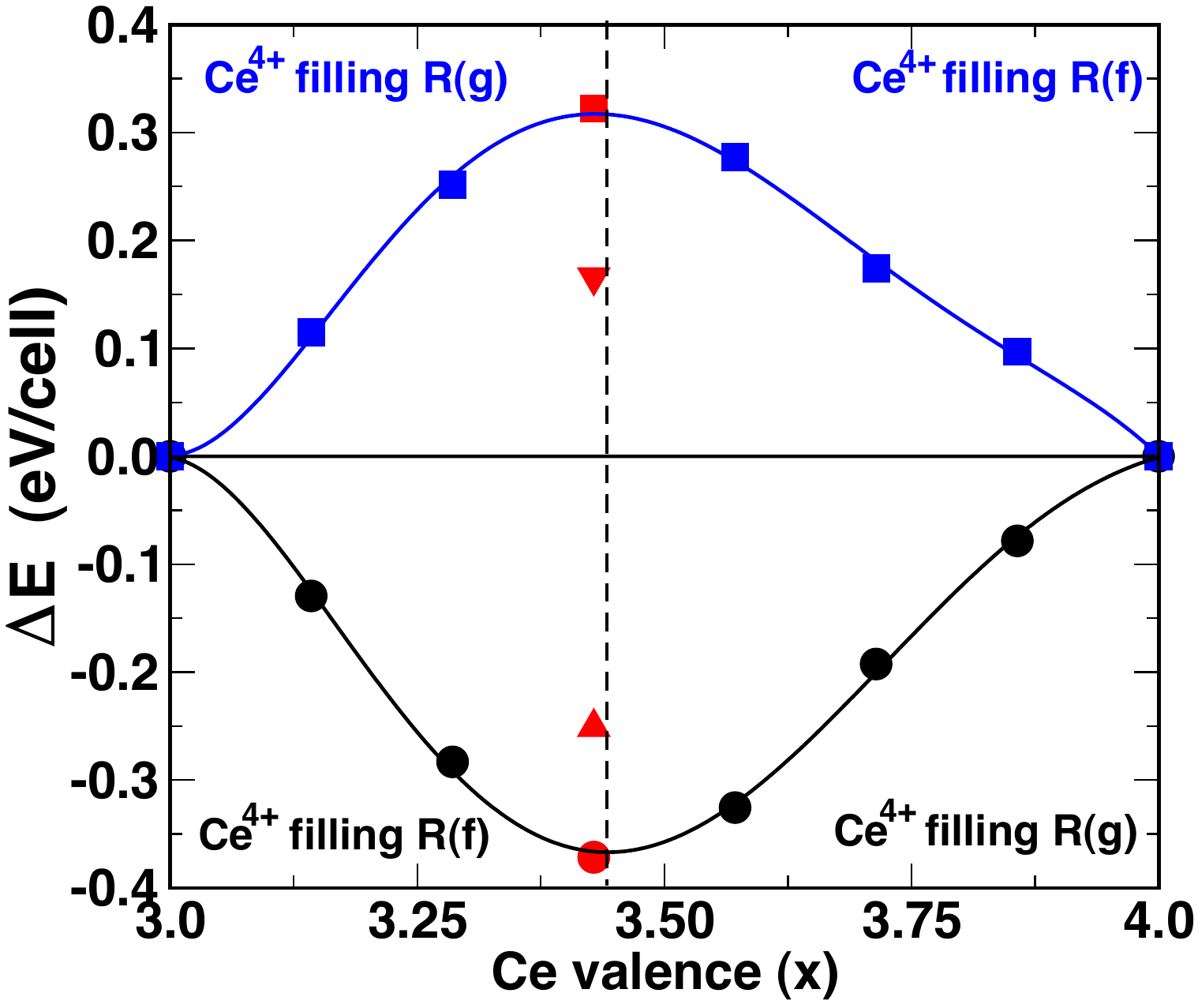}
\caption {(Color online) Similar to Fig.~\ref{fig2} but for (Ce$_{87.5}$La$_{12.5}$)$_2$Fe$_{14}$B. Here La is substituted at its favored $4$g-site and Ce$^{3+}$ and Ce$^{4+}$ are distributed over the remaining $7$ RE-sites in all possible ways.} 
\label{fig3}
\end{figure}

{\par}To improve magnetic properties requires engineering Ce- or Fe-sites in such a way to push the Ce-valency either towards $3+$ or $4+$. 
Due to the dependence of Ce-valence on the steric volume,\cite{Alam13} one way to manipulate the Ce valency is to vary the unit cell volume by forming pseudo-ternary compounds. 
We studied two compounds, i.e., (Ce,La)$_2$Fe$_{14}$B and Ce$_2$(Fe,Co)$_{14}$B; the former (later) should increase (reduce) the unit cell volume.
First, we dope RE Ce-sites by La in Ce$_{2}$Fe$_{14}$B. 
Out of the two inequivalent RE sites $4$f and $4$g, La (being larger than Ce) prefers to occupy the $4$g sites. Figure \ref{fig3} shows $\Delta$E vs. Ce valence with $12.5\%$  of La doping (1 out of 8 sites) in Ce$_{2}$Fe$_{14}$B.  
From the data in the vicinity of the minimum ($x\sim3.5$), La-doping clearly moves the Ce-valency towards $3+$ relative to the undoped case.
 This effect is in accord with the steric volume argument: a La-ion, being larger than Ce, expands the lattice when doped in Ce$_{2}$Fe$_{14}$B,  enhancing the steric volume of Ce site(s) and supports a more trivalent-like state. 
Steric volume is an important factor controlling the Ce chemical valence, as also evidenced in hydrogenated Ce$_{2}$Fe$_{14}$B and Ce$_{2}$Fe$_{17}$ compounds.\cite{Capehart93,Fruchart87}

\begin{figure}[t]
\centering
\includegraphics[width=6.0cm]{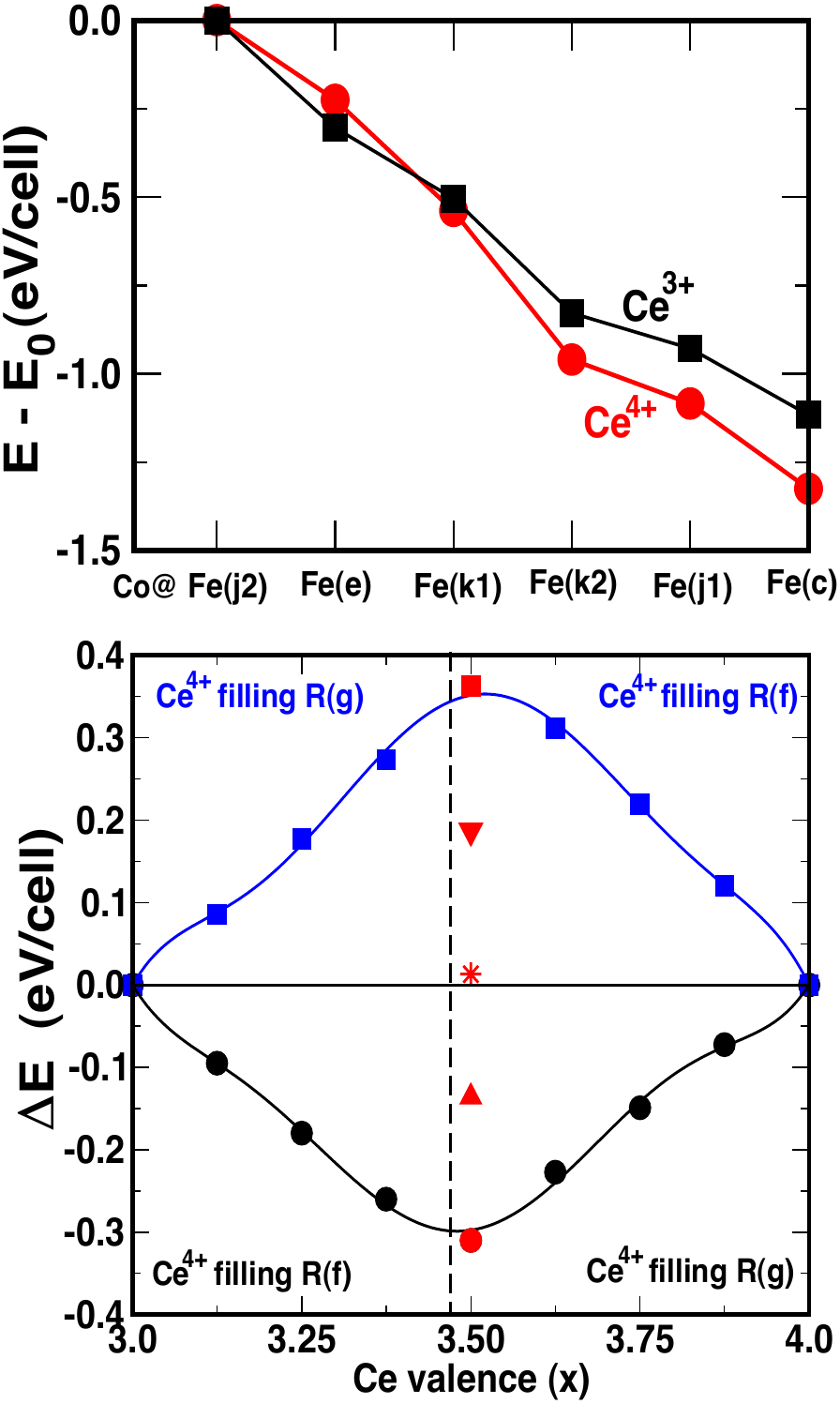}
\caption {(Color online) Site-preference energy for Co doping at various Fe-sites in Ce$_2$Fe$_{14}$B (top). E$_0$ is a reference when Co is on the Fe(j2) site. (Bottom) Same as Fig.~\ref{fig2}, but for Ce$_2$(Fe$_{13}$Co)B with Co occupying Fe(c) sites.} 
\label{fig4}
\end{figure}

{\par} Next, we study the effects of Co doping on Fe-sites in Ce$_{2}$Fe$_{14}$B, which is known that to enhance the Curie temperature and magnetic anisotropy and, hence, a reason for our choice. 
Ce$_{2}$(Fe,Co)$_{14}$B crystallizes in the same P$4_2$/mnm structure as Ce$_{2}$Fe$_{14}$B, which has six
inequivalent Fe-sites,\cite{Alam13} i.e., Fe(k1), Fe(k2), Fe(j1), Fe(j2), Fe(e) and Fe(c).
First we verified the Co site preference on these symmetry distinct Fe-sites. 
Figure \ref{fig4} (top) shows the site-preference energy for both Ce$^{3+}$ and Ce$^{4+}$ when $1$ out of $14$ Fe-sites are doped with a Co atom.
Results indicate that Fe(j2) site has the strongest preference for not occupying Co, as it costs the highest
energy. 

{\par}This finding is supported by two arguments:
(1) Co and Fe differ a little in size (R$_{\text{Co}} <$ R$_{\text{Fe}}$); out of $6$ Fe-sites in 2-14-1, Fe(j2) has the largest coordination volume -- a reason for Co to avoid j2-sites. 
(2) j2-sites in 2-14-1 and C(dumbbell)-sites in rhombohedral Ce$_{2}$Fe$_{17}$ structure\cite{Herbst91} are reported to be crystallographically, as well as magnetically,  cognates.\cite{Herbst86} 
In other words, each of these sites have the largest number of near-neighbor Fe-ions and the largest moment. 
Also, these sites are the only transition-metal sites that have a major ligand line -- perhaps another reason behind the unfavorability of Co to occupy j2-sites.
Fe(c) and Fe(j1) sites have the strongest preference towards Co. 
This site preference can be justified from the large affinity of Co towards rare-earth, i.e., those
transition-metal-sites that acquire the highest RE-coordination will prefer to have Co on it. 
Fe(c) and Fe(j1) indeed has the highest coordination of RE around it.

{\par} Figure \ref{fig4}shows $\Delta$E~vs.~Ce valence for Ce$_{2}$(Fe$_{13}$Co)B with Co doped on the energetically most favorable Fe(c) site.
Unlike  Ce$_{2}$Fe$_{14}$B (Fig.~\ref{fig2}), $7.14\%$ Co doping ($1$ out of $14$) already favors the mixing of Ce valence at the Ce$^{4+}$ end, i.e., no positive (unfavorable) $\Delta$E. 
Again, compared to the Ce valence in Ce$_{2}$Fe$_{14}$B, Co doping pushes the valency of Ce towards $3+$ (similar 
argument about the location of minima holds in this case).
This, however, does not jibe with the volume argument used for La doping.
Co being smaller than Fe leads to a volume reduction that should move the Co valence towards $4+$ via steric volume, instead of $3+$. 
In this case, the local chemistry and the associated local steric volume of RE-sites plays an important role in determining the Ce valency than the simple concept based on global volume reduction. 

\begin{figure}[t]
\centering
\includegraphics[width=9.0cm]{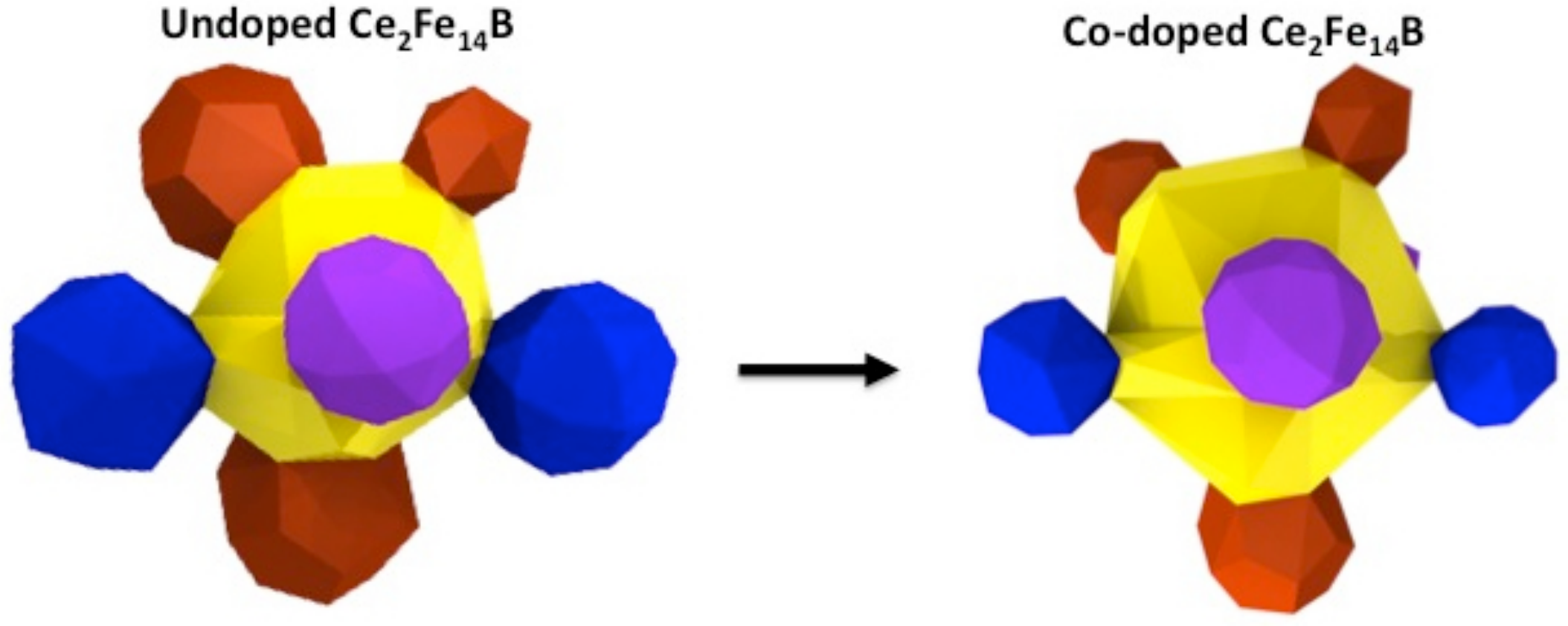}
\caption {(Color online) Expansion of the steric [Voronoi] volume of the central Ce-site (yellow) after Co-doping. Inequivalent Fe sites are denoted by blue, red, and purple.  With Co-doping, the steric volume of the Ce site as enlarged, at the expense of the Fe sites and associated charge.} 
\label{fig5}
\end{figure}

{\par}It is well known that Co has a strongly affinity to RE elements (Ce in this case), as such Ce favors a high  coordination number of Co.
In  2-14-1, Fe(c) sites with 4 Ce and Fe(j1) sites with 3 Ce have the highest number of RE neighbors. 
These Fe-sites are indeed the energetically most favorable site for Co, see Fig.~\ref{fig4} (top). 
Now, because R$_{\text{Co}}<$~R$_{\text{Fe}}$, the accumulation of a large number of Co around Ce-site causes the formation of major ligands, given by lines connecting faces of Voronoi polyhedra, allowing more room and an expansion of the local Voronoi volume around the Ce-site. (These Voronoi polyhedra and volumes were determined by inscribed radii given by saddle-points in the electronic density.\cite{Alam11}) 
Note how the central Ce-polyhedra expands (Fig.~\ref{fig5}) due to Co-doping on Fe-sites. 
Thus, although Co doping reduces the unit cell volume, the local steric volume around the Ce-site is enhanced which shifts the Ce valency towards $3+$.
This phenomenon is based on the local chemistry and the nature of hybridization of Ce-ion with its neighboring atoms, instead of the simple volume argument alone.

{\par}In summary, we have presented an ``alloy'' approach to predict reliably the mixed-valency properties in complex compounds, for which DFT+U methods are essential, and reveal the electronic origin for such behavior. 
For Ce-based materials, cerium does not have a well-defined valence; rather its {\it f}-electrons fluctuate between two extreme valence states dependent upon local atomic configurations (site occupancy) with a distributions of $3+$ and $4+$ Ce.
The energy difference between these states/configurations is a few $me$V-atom$^{-1}$, so associated anomalies are observed, $e.g.$, under pressure. 
The mechanism for such a transition and the reason for differing valence states was not yet well understood. 
Doping puts the material under an chemical pressure -- La doping at Ce sites expands the lattice (as expect from steric volume arguments), while a transition-metal dopant like Co at Fe sites shrinks it (in contrast to steric volume arguments).
Here we predicted the mixed valency of Ce in pure Ce and Ce$_2$Fe$_{14}$B, in agreement with experiment; then, we addressed two different types of doping (La at Ce-sites and Co at Fe-sites) to reveal how both steric volume and local chemistry influence the Ce valence in compounds, reflected in the nature of hybridization with neighboring atoms, i.e., Ce site preferences arising from the large electron affinity of Co towards rare-earth elements.

\vspace{0.1cm}
Work at Ames Laboratory was supported by the U.S. Department of Energy (DOE) ARPA-E REACT Program (contract 0472-1526), using capabilities maintained and supported by the Office of Basic Energy Sciences in our Division of Materials Science and Engineering. The Ames Laboratory is operated for the U.S. DOE by Iowa State University under contract DE-AC02-07CH11358.
\vspace{-0.5cm}

\end{document}